\documentclass[epj]{webofc}
\usepackage[utf8]{inputenc}
\usepackage[varg]{txfonts}   
\usepackage{booktabs}
\usepackage{xcolor}
\usepackage{float}
\usepackage{dsfont}
\definecolor{darkred}{rgb}{0.4,0.0,0.0}
\definecolor{darkgreen}{rgb}{0.0,0.4,0.0}
\definecolor{darkblue}{rgb}{0.0,0.0,0.4}
\usepackage[bookmarks,linktocpage,colorlinks,
    linkcolor = darkred,
    urlcolor  = darkblue,
    citecolor = darkgreen]{hyperref}
\usepackage{subcaption}
\wocname{EPJ Web of Conferences}
\woctitle{Lattice2017}
\begin{document}
%
\selectlanguage{english}
\title{%
Worldlines and worldsheets for non-abelian lattice field theories: Abelian color fluxes and Abelian color cycles
}
\author{%
\firstname{Christof}  \lastname{Gattringer}\thanks{Speaker, \email{christof.gattringer@uni-graz.at}},
\firstname{Daniel}  \lastname{G\"oschl}\thanks{\email{daniel.goeschl@uni-graz.at}},
\firstname{Carlotta}  \lastname{Marchis}\thanks{Speaker, \email{carlotta.marchis@uni-graz.at}}
}
\institute{%
Institut f\"ur Physik, Universit\"at Graz, 8010 Graz, Austria 
}
\abstract{%
We discuss recent developments for exact reformulations of lattice field theories in terms of worldlines and worldsheets. In particular we 
focus on a strategy which is applicable also to non-abelian theories: traces and matrix/vector 
products are written as explicit sums over color 
indices and a dual variable is introduced for each individual term. These dual variables correspond to fluxes in both, space-time 
and color for matter fields (Abelian color fluxes), or to fluxes in color space around space-time plaquettes for gauge fields
(Abelian color cycles). Subsequently all original degrees of freedom, i.e., matter fields and gauge links, can be integrated out.
Integrating over complex phases of matter fields gives rise to constraints that enforce conservation of matter flux on all sites. Integrating out
phases of gauge fields enforces vanishing combined flux of matter- and gauge degrees of freedom. The constraints give rise to a system of 
worldlines and worldsheets. Integrating over the factors that are not phases (e.g., radial degrees of freedom or contributions from 
the Haar measure) generates additional weight factors that together with the constraints implement the full symmetry of the conventional
formulation, now in the language of worldlines and worldsheets. We discuss the Abelian color flux and Abelian color cycle strategies for
three examples: the SU(2) principal chiral model with chemical potential coupled to two of the Noether charges, SU(2) 
lattice gauge theory coupled to staggered fermions, as well as full lattice QCD with staggered fermions. For the principal chiral model we 
present some simulation results that illustrate properties of the worldline dynamics at finite chemical potentials.

}
\maketitle
\section{Introduction}\label{intro}

Dualization techniques have recently received quite some attention in the lattice community as a possible way to solve complex 
action problems caused by finite density or topological terms (see, e.g., the reviews 
\cite{Chandrasekharan:2008gp,deForcrand:2010ys,Wolff:2010zu,Gattringer:2014nxa,Gattringer:2016kco}). It is now well established 
that essentially all abelian bosonic theories with chemical potential can be exactly mapped to dual representations where the partition
function is a sum over worldlines and worldsheets, and all weights are real and positive such that Monte Carlo 
simulations can be done directly in terms of the new degrees of freedom. 

For non-abelian theories the program of dualizing the system in terms of worldlines and worldsheets is not very advanced yet. The 
main reason is the problem of reordering the non-abelian variables of the conventional representation after expansion of the local 
Boltzmann factor, such that they can be integrated out in closed form. In particular for non-abelian gauge theories the lack of a 
suitable reordering strategy quickly leads to a proliferation of rather non-local couplings in the resulting dual representation. However, 
for several (spin-) systems with non-abelian symmetries successful complete dualizations\footnote{Complete dualizations
in the sense that the full theory 
was dualized and not only leading terms of a strong coupling series -- see however the discussion below.} were discussed recently 
\cite{Chandrasekharan:2008gp,Bruckmann:2014sla,Bruckmann:2015sua,Bruckmann:2015hua,Bruckmann:2015uhd,Bruckmann:2016hes,Bruckmann:2016txt,Bruckmann:2016fuj,Vairinhos:2014uxa,Vairinhos:2015ewa,Rindlisbacher:2015xku,Rindlisbacher:2016cpj,Rindlisbacher:2017dph,Wolff:2009kp,Wolff:2010qz}. In all cases a suitable representation was found such that after 
strong coupling expansion of local Boltzmann factors the original degrees of freedom could be integrated out in closed form. 
This is a strategy that led to interesting worldline representations for several non-abelian systems and is also the basis for the approach
reviewed here.

In our contribution we discuss results for a recently introduced strategy for the dualization of non-abelian theories. In the so-called
Abelian color flux and Abelian color cycle approaches one writes all traces and matrix/vector products as sums over color indices 
(or more generally over 'internal indices') and introduces a dual variable for each individual contribution. At this stage of the 
dualization the dual variables are simply the expansion indices used for the Taylor series of the individual Boltzmann factors. 
All terms in the corresponding expansion are complex numbers such that the reordering problem is solved. Using a suitable 
representation all original degrees of freedom can be integrated out. This leads to a set of constraints that come from integrating out 
the phases of the original fields. As a result the dual variables are now restricted to a system of worldlines and worldsheets. Integrating 
the degrees of freedom that are not phases, such as radial degrees of freedom or Haar-measure contributions, leads to weight factors 
that together with the constraints implement the symmetries of the conventional formulation on the worldline/worldsheet representation.  

In order to develop the Abelian color flux and cycle concepts in this paper we discuss them for three different systems: the SU(2) 
principal chiral model with chemical potentials coupled to two of the conserved charges, the SU(2) lattice gauge theory coupled 
to staggered fermions, as well as full QCD with staggered fermions. 

\section{The principal chiral model}

The principal chiral model provides a simple example for developing the dualization strategy with Abelian color fluxes outlined in the 
introduction. It has the advantage that the structure of the action is simpler than in a gauge theory -- it consists only of nearest 
neighbor terms -- and thus the idea of the Abelian color flux dualization \cite{Gattringer:2017hhn}
can be presented in a more transparent way.  We couple 
chemical potentials $\mu_1, \mu_2$  
to two of the conserved charges, which introduces a complex action problem in the conventional approach that 
is overcome in the dual representation. Furthermore the chemical potentials allow one to monitor how the conserved charges are manifest 
in the dual representation -- they turn out to correspond to temporal winding numbers for two species of worldlines. 

\subsection{Dualization with Abelian color fluxes}

The conventional degrees of freedom of the model are SU(2) matrices $U_x$ assigned to the sites $x$ of a 
$d$-dimensional $N^{d-1} \times N_t$ lattice with periodic boundary conditions. The action sums traced nearest neighbor
terms of the $U_x$ over all links $x,\nu$ of the lattice,
\begin{equation}
S  \; = \;  - \frac{J}{2}  \sum_{x, \nu} \left( 
\textnormal{Tr}  \left[ e^{\, \delta_{\nu,d} \sigma_3 \frac{\mu_1 + \mu_2}{2}} \, U_x \, 
e^{\, \delta_{\nu,d} \sigma_3 \frac{\mu_1 - \mu_2}{2}} \, U_{x+\hat{\nu}}^\dagger \right] +  
\textnormal{Tr}  \left[ e^{- \delta_{\nu,d} \sigma_3 \frac{\mu_1 - \mu_2}{2}} \, U_x^\dagger \,
e^{- \delta_{\nu,d} \sigma_3 \frac{\mu_1 + \mu_2}{2}} \, U_{x+\hat{\nu}} \right] \, \right) \;   .
\label{pcm_latticeaction}
\end{equation}
The two chemical potentials $\mu_1$ and $\mu_2$ give different weights to hops in the forward and backward 
temporal direction (the direction $\nu = d$). Obviously $S$ has an imaginary part for finite $\mu_\lambda$, $\lambda = 1,2$, 
giving rise to a complex action problem in the conventional form. To obtain the partition function of the model we 
integrate the Boltzmann factor with a product of Haar measures 
$\int \! DU = \prod_x \int_{SU(2)} \textnormal{d} U_x$ and the partition sum is given by $Z = \int \! D[U] \, e^{-S}$.

As already outlined in the introduction, for the Abelian flux representation we write all traces and matrix products as explicit sums over color
indices\footnote{Although the site variables $U_x \in$ SU(2) have nothing to do with gluons -- still we refer to their
indices as ''color indices''.}. After some reshuffling of terms the action in the exponent of the Boltzmann factor is written as a double 
sum over color indices, giving rise to \cite{Gattringer:2017hhn}
\begin{equation}
Z \; = \; \int \!\!  D[U] \exp \left(  J  \sum_{x,\nu} \sum_{a,b = 1}^2 M_\nu^{ab} U_x^{ab}  {U_{x+\hat{\nu}}^{ab}}^{\!\!\! \star} \right)
\; = \; \int \!\! D[U] \prod_{x,\nu} \prod_{a,b = 1}^2 \, e^{\, J M_\nu^{ab} U_x^{ab}  {U_{x+\hat{\nu}}^{ab}}^{\!\!\! \star} }\; ,
\label{pcm_trafo1}
\end{equation}
where we have taken into account the $\mu_\lambda$-dependence in the factors $M_\nu^{ab}$ defined as 
\begin{equation}
M_ \nu^{11} = e^{\, \mu_1 \delta_{\nu,d} } \quad , \quad M_ \nu^{22} = e^{\, -\mu_1 \delta_{\nu,d} } \quad , \quad
M_ \nu^{12} = e^{\, \mu_2 \delta_{\nu,d} } \quad , \quad M_ \nu^{21} = e^{\, -\mu_2 \delta_{\nu,d}} \; .
\end{equation}
In the second step of (\ref{pcm_trafo1}) we have converted the sum in the exponent into a product over Boltzmann factors
for the individual terms. Note that in these individual Boltzmann factors only products of complex numbers appear which will allow
us to reorder the contributions and organize them according to the sites $x$ where we integrate over the conventional 
degrees of freedom $U_x^{ab}$. In order to access individual entries $U_x^{ab}, \, a,b = 1,2$ we use the following explicit 
representation of the SU(2) matrices and the Haar measure
($\theta_x \in [0,\pi/2], \, \alpha_x \in [-\pi, \pi], \, \beta_x \in [-\pi, \pi] $)
\begin{equation}
U_x \; = \; \left[ \begin{array}{cc} 
\cos \theta_x e^{\, i \alpha_x} & \sin \theta_x e^{\, i \beta_x} \\
-\sin \theta_x e^{\, - i \beta_x} & \cos \theta_x e^{\, - i \alpha_x} 
\end{array} \right] \quad \mbox{with} \quad \textnormal{d}U_x \; = \; 2 \sin \theta_x \, \cos \theta_x \, \textnormal{d} \theta_x \, 
\frac{\textnormal{d} \alpha_x}{2\pi} \frac{\textnormal{d} \beta_x}{2\pi} \; .
\label{pcm_parameterization} 
\end{equation}

The next step is to expand the individual Boltzmann factors for each combination of $x, \nu, a, b$ into a power series and to 
reorder the factors according to the sites $x$. The partition sum assumes the form 
\begin{eqnarray}
\hspace*{-20mm}
Z \!\! &\!\!\!\! = \!\!\!\!\!& \!\! \int \!\!\! D[U] \prod_{x,\nu} \prod_{a,b} \sum_{j_{x,\nu}^{\, ab} = 0}^\infty 
\frac{ ( J M_\nu^{ab} )^{j_{x,\nu}^{\,ab}}}{j_{x,\nu}^{\,ab} \, !} \, \left[ U_x^{ab}  {U_{x+\hat{\nu}}^{ab}}^{\!\!\! \star} \right]^{j_{x,\nu}^{\,ab}} 
 =  \sum_{\{ j \}} \!\!W_{J,\mu}[j] \!\! \int \!\!\! D[U] \prod_{x,\nu} \prod_{a,b} 
\big[ U_x^{ab} \big]^{j_{x,\nu}^{\,ab}}   \big[{U_{x}^{ab}}^{\star} \big]^{j_{x-\hat{\nu},\nu}^{\,ab}}
\nonumber \\ 
&\!\! = \!\!\!& \sum_{\{ j \}} W_{J,\mu}[j] \, \prod_x 2 \!\! \int_0^{\frac{\pi}{2}} \!\! \textnormal{d}\theta_x \; 
(\cos \theta_x)^{1 + \sum_\nu\big[ j_{x,\nu}^{\,11} + j_{x,\nu}^{\,22} + j_{x-\hat{\nu},\nu}^{\,11} + j_{x-\hat{\nu},\nu}^{\,22}\big]} \,
(\sin \theta_x)^{1 + \sum_\nu\big[ j_{x,\nu}^{\,12} + j_{x,\nu}^{\,21} + j_{x-\hat{\nu},\nu}^{\,12} + j_{x-\hat{\nu},\nu}^{\,21}\big]}
\nonumber \\
&& \hspace{5mm} \times \int_{-\pi}^{\pi} \frac{\textnormal{d}\alpha_x}{2\pi} \,
e^{\, i \alpha_x \sum_\nu\big[ \big( j_{x,\nu}^{\,11} - j_{x,\nu}^{\,22}\big) - \big(j_{x-\hat{\nu},\nu}^{\,11} - j_{x-\hat{\nu},\nu}^{\,22}\big)\big]}
 \int_{-\pi}^{\pi} \frac{\textnormal{d}\beta_x}{2\pi} \,
e^{\, i \beta_x \sum_\nu\big[ \big( j_{x,\nu}^{\,12} - j_{x,\nu}^{\,21}\big) - \big(j_{x-\hat{\nu},\nu}^{\,12} - j_{x-\hat{\nu},\nu}^{\,21}\big)\big]} \; .
\label{pcm_trafo2}
\end{eqnarray}
For each individual Boltzmann factor a summation variable $j_{x,\nu}^{\,ab} \in \mathds{N}_0$ was introduced and $\sum_{\{ j \}} $
denotes the sum over all configurations of these variables. The weight factor $W_{J,\mu}[j]$ collects all weights generated in the
expansion of the Boltzmann factors, 
\begin{equation}
W_{J,\mu}[j] \, \equiv \, \prod_{x,\nu} \prod_{a,b} \frac{ ( J M_\nu^{ab} )^{j_{x,\nu}^{\,ab}}}{j_{x,\nu}^{\,ab} \, !} \; = \; 
e^{\, \mu_1 \! \sum_x \! \big[ j_{x,d}^{\,11} - j_{x,d}^{\,22} \big]}
e^{\, \mu_2 \! \sum_x \! \big[ j_{x,d}^{\,12} - j_{x,d}^{\,21} \big]}
\prod_{x,\nu} \prod_{a,b} \frac{J^{j_{x,\nu}^{\,ab}}}{j_{x,\nu}^{\,ab} \, !} \; .
\end{equation}
In the last step of (\ref{pcm_trafo2}) we have reordered the factors and collected the terms $U_x^{ab}$ for all combinations of $x, a, b$. 
Subsequently the explicit representation (\ref{pcm_parameterization}) was inserted giving rise to integrals over $\theta_x$, $\alpha_x$ and 
$\beta_x$ at each site $x$. The integrals over $\alpha_x$ and $\beta_x$ are Kronecker deltas for the integer valued combinations 
of the $j_{x,\nu}^{\,ab}$ in the respective exponents, and constitute constraints for these combinations at each site $x$. 
The integrals over $\theta_x$ give rise to beta-functions that can be simplified as fractions of factorials,
since the constraints imply that the exponents of $\cos \theta_x$ and $\sin \theta_x$ are odd.

The worldline representation can be simplified further by introducing new variables: the ''flux variables''
$k_{x,\nu}^\lambda \in \mathds{Z}, \lambda = 1,2$ and the ''auxiliary variables'' 
$m_{x,\nu}^\lambda \in \mathds{N}_0, \lambda = 1,2$. They are defined as
\begin{equation}
k_{x,\nu}^{1} = j_{x,\nu}^{\,11} - j_{x,\nu}^{\,22} \; , \quad 
k_{x,\nu}^{2} = j_{x,\nu}^{\,12} - j_{x,\nu}^{\,21} \; , \quad
m_{x,\nu}^{1} = \frac{j_{x,\nu}^{\,11} + j_{x,\nu}^{\,22} - |k_{x,\nu}^1|}{2}  \; , \quad 
m_{x,\nu}^{2} = \frac{j_{x,\nu}^{\,12} + j_{x,\nu}^{\,21} - |k_{x,\nu}^2|}{2} \; .
\end{equation}  
Expressing the $j_{x\nu}^{\,  ab}$ in terms of the $k_{x,\nu}^\lambda$ and $m_{x,\nu}^\lambda$ one arrives at the final form for the 
worldline representation of the partition function, which we write as
\begin{equation}
Z \; = \; \sum_{\{k,m\}} W_J[k,m] \, W_H[k,m] \, W_\mu[k] \; 
\prod_x \prod_{\lambda = 1}^2 \, \delta \left( {\nabla} {k}^{\, \lambda}_x \right) \; .
\label{pcm_worldlineZ}
\end{equation}
The partition function is a sum $\sum_{\{k,m\}}$ over all possible configurations of the flux variables 
$k_{x,\nu}^\lambda \in \mathds{Z}, \lambda = 1,2$ and the auxiliary variables 
$m_{x,\nu}^\lambda \in \mathds{N}_0, \lambda = 1,2$, which are both assigned to the links of
the lattice. Only the flux variables are subject to constraints, which are implemented as product of Kronecker deltas 
(we use the notation $\delta(n) \equiv \delta_{n,0}$). They enforce a zero divergence condition for both $k_{x,\nu}^1$ 
and $k_{x,\nu}^2$ at every site $x$, where the discretized divergence is defined as 
${\nabla} {k}^\lambda_x \equiv \sum_\nu [ k_{x,\nu}^\lambda - k_{x - \hat{\nu},\nu}^\lambda ]$. A vanishing divergence 
implies that at each site $x$ the total flux of $k_{x,\nu}^\lambda$ has to vanish and admissible configurations of the fluxes of 
$k_{x,\nu}^\lambda, \lambda = 1,2$ are closed worldlines for the two species of fluxes. 

The admissible configurations of the $k$- and $m$-variables come with weights which we split into three
factors as follows: $W_J[k,m]$ collects the factors that were generated when expanding the individual Boltzmann terms.
The weight factor $W_H[k,m]$ contains the contributions from the $\theta_x$-integrations, which gives rise to the beta functions
that here simplify to fractions of factorials. Finally $W_\mu[k]$ is the term that couples the fluxes to the chemical potentials. The first two weight factors are 
\begin{equation} 
W_J[k,m] \, = \, \prod_{x,\nu} \prod_{\lambda = 1}^2 
\frac{ J^{\, D_{x,\nu}^\lambda}}{( D_{x,\nu}^\lambda - m_{x,\nu}^\lambda)! \, m_{x,\nu}^\lambda !}
 \; , \;
W_H[k,m] \, = \, \prod_{x} \frac{ \prod_{\lambda = 1}^2 \left( \frac{1}{2}
\sum_\nu [D_{x,\nu}^\lambda + D_{x-\hat{\nu},\nu}^\lambda]\right)!}
{\left( 1+ \frac{1}{2}\sum_\nu \sum_\lambda [D_{x,\nu}^\lambda + D_{x-\hat{\nu},\nu}^\lambda]\right)!} \; ,
\label{WJWH}
\end{equation}
where we use the abbreviation $D_{x,\nu}^\lambda \; \equiv \; |k_{x,\nu}^\lambda| + 2 m_{x,\nu}^\lambda$. The weight factor 
$W_\mu[k]$ is given by
\begin{equation}
W_\mu[k] \; = \; \prod_{\lambda = 1}^2 \prod_x e^{\, \mu_\lambda \, k_{x,d}^\lambda} 
\; = \; \prod_{\lambda = 1}^2e^{\, \mu_\lambda \sum_x k_{x,d}^\lambda} \; = \; 
e^{ \, \mu_1 \beta \, \omega_1[k] } \, e^{ \, \mu_2 \beta \, \omega_2[k] } \; .
\label{Wmu}
\end{equation}
In the last step the identity $\sum_x k_{x,d}^\lambda = N_t \, \omega_\lambda [k]$ was used,
where $\omega_\lambda [k], \lambda = 1,2$ are the temporal net winding numbers of the two species of fluxes 
described by the two sets of dual variables $k^\lambda_{x,d}, \, \lambda = 1,2$.  The 
temporal extent $N_t$ is the inverse temperature in lattice units and on the right hand side of (\ref{Wmu})
we replaced $N_t$ by the more usual symbol $\beta$.

\begin{figure}
\begin{center}
\includegraphics[scale=0.5,clip]{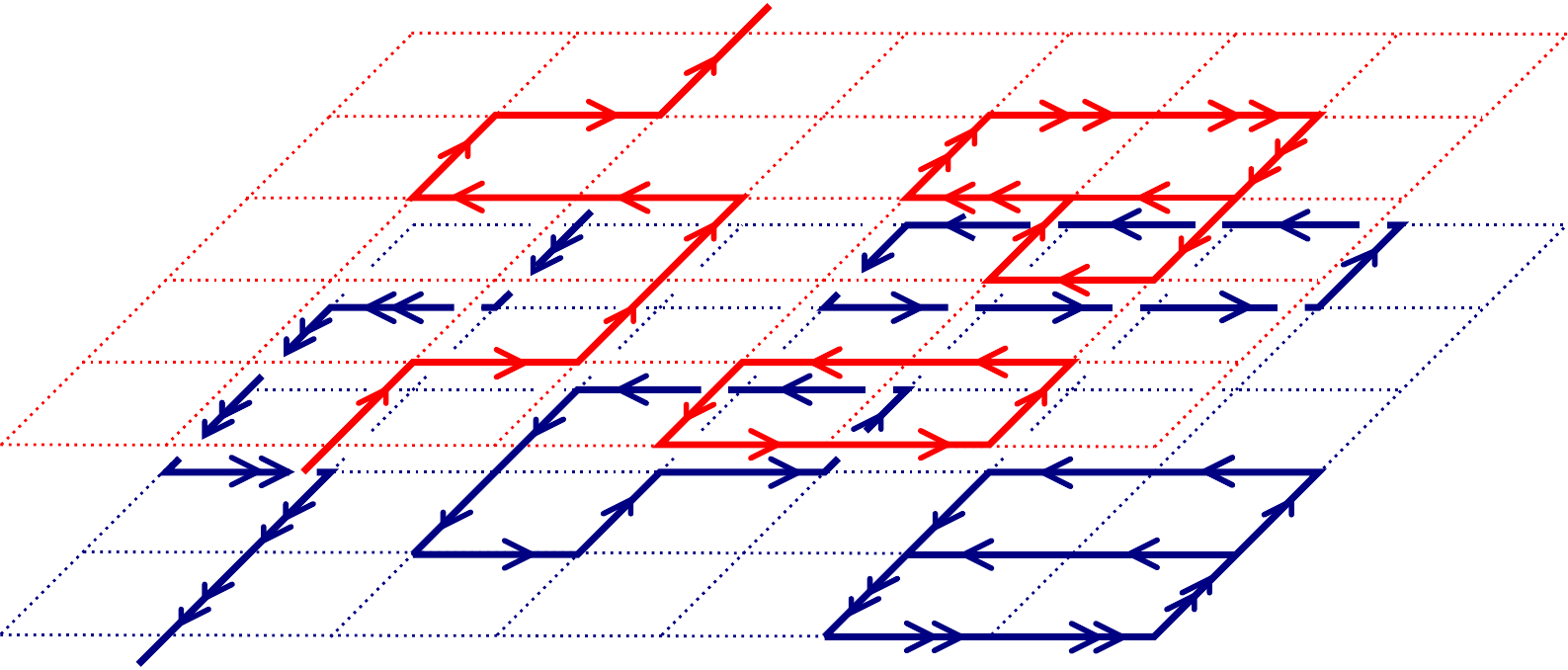}
\end{center}
\caption{Illustration of the worldline form of the principal chiral model: Admissible configurations consist of two species of worldlines
with individually conserved fluxes, which we here illustrate using a space-time lattice (2-dimensional in our illustration) with two
layers for the two species of fluxes. On the sites of the lattice the two species of worldlines interact via weight factors that come from 
the Haar measure integrals and implement the original SU(2) symmetry in the worldline representation.}
\label{fig:worldlines}	
\end{figure}

Let us discuss some properties of the resulting worldline representation: It is obvious that all weight factors are real and positive 
also for finite $\mu_\lambda$. Consequently in the worldline representation (\ref{pcm_worldlineZ}) the complex action problem is solved
and a Monte Carlo simulation is possible in terms of the dual variables. As we have discussed in the previous paragraph, the 
chemical potentials couple to topological quantities in the dual representation, the temporal winding numbers of the two species
of worldlines. This property can also be viewed as a guiding principle for the construction of worldline 
representations\footnote{Note that a worldline representation is not necessarily unique -- for a different worldline representation 
of the principal chiral model see \cite{Rindlisbacher:2015xku}.}: Identify a set of Noether charges in the conventional representation and 
choose a parameterization such that the corresponding chemical potentials appear with phases, such that after integrating 
over the conventional degrees of freedom, the chemical potentials couple to temporal winding numbers of worldlines.   

The geometrical interpretation of the net-particle numbers as temporal winding numbers implies that for each individual 
configuration of the dual variables we can determine the net-particle numbers
as integers -- a property which is not shared by the conventional representation where the net particle number corresponds to a 
lattice discretization of the continuum Noether charge, which is not necessarily integer. Thus, in the worldline representation it is 
straightforward to set up a canonical simulation, i.e., a simulation at fixed net-particle number (compare \cite{Orasch:2017niz}). 

Let us finally discuss how the worldline representation implements the SU(2) symmetry of the original representation: The worldlines for 
the two species of conserved fluxes would correspond to two independent U(1) charges. However, at each site $x$ of the lattice
the two species of worldlines interact with each other and the auxiliary variables via the weight factors $W_J[k,m]$. These weights
come from integrating over the contributions containing the angles $\theta_x$ with the Haar measure, thus implementing the SU(2)
symmetry of the conventional representation. A graphical illustration of the emerging geometrical picture is shown in Fig.~\ref{fig:worldlines}.

\subsection{Monte Carlo simulation and some results}

\begin{figure}[t]
\begin{center}
\includegraphics[scale=0.35,clip]{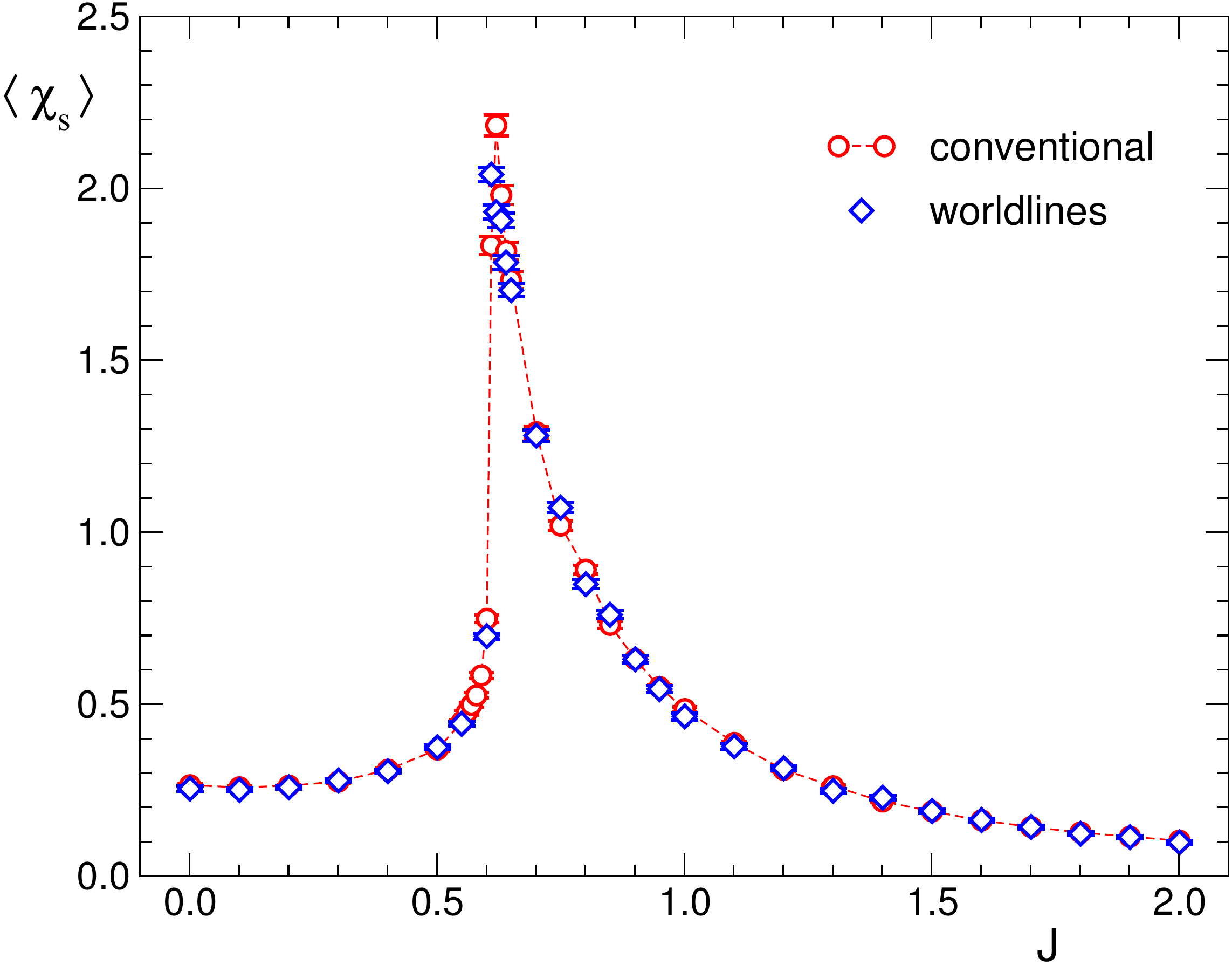}
\end{center}
\caption{We compare the results for the susceptibility $\chi_s$ from the conventional- (circles) and the worldline (diamonds) simulations
($16^4, \, \mu_1 = \mu_2 = 0$).}
\label{fig:chiUcompare}	
\end{figure}

The dual representation of the principal chiral model (\ref{pcm_worldlineZ}) does not exhibit a sign problem at finite $\mu_\lambda$
and a Monte Carlo simulation is possible in terms of the worldlines. In this brief explorative 4-d study we consider bulk observables, 
which can be obtained as derivatives of the logarithm of the partition sum. More specifically, we investigate the action density 
$\langle s \rangle$ and the particle number densities $\langle n_{\lambda} \rangle$, as well as the action density 
susceptibility $\chi_s$ and the particle number susceptibilities $\chi_{n_{\lambda}}$ defined as
\begin{equation}
 \langle s\rangle = \frac{1}{8V} \frac{\partial}{\partial J} \ln(Z) \ , \quad
 \chi_s = \frac{1}{8V} \frac{\partial^2}{\partial J^2} \ln(Z) \ , \quad 
 \langle n_{\lambda} \rangle = 
 \frac{1}{V} \frac{\partial}{\partial\mu_\lambda} \ln(Z) \ ,  \quad
 \chi_{n_{\lambda}} = \frac{1}{V} \frac{\partial^2}{\partial{\mu_\lambda}^2} \ln(Z) \ .
\end{equation}
These derivatives can be evaluated also for the dual form and the observables read 
\begin{align}
\label{equ:observables_worldline}
\langle s \rangle = \frac{1}{4 V J} \left\langle \bar{D}\right\rangle ,\,
\chi_s = \frac{1}{4 V J} 
\left[ \left\langle  \bar{D}^2 \right\rangle \! -\! \left\langle  \bar{D} \right\rangle^2 \!\!-\! \left\langle  \bar{D} \right\rangle \right] , \,
\langle n_{\lambda} \rangle = \frac{1}{N_{s}^{3}} \Bigl\langle \omega_\lambda [k] \Bigr\rangle, \,
\chi_{n_{\lambda}} = \frac{1}{N_{s}^{3}} \left[ \left\langle  {\omega_\lambda [k]}^2 \right\rangle - \Bigl\langle \omega_\lambda [k] \Bigr\rangle^2 \right],
\end{align}
where we use the shorthand notation $\bar{D} = \sum_{x,\nu} \sum_{\lambda} D_{x,\nu}^{\lambda}$.

Our strategy for generating new configurations in the Monte Carlo simulation combines a local Metropolis update for the unconstrained 
$m$-variables and two separate updates for the $k$-variables which have to obey the constraint of vanishing divergence. 
For the latter we perform local changes where we offer to change the $k$-flux on the four links of a plaquette by one unit.
In addition we explicitly insert winding loops of $k$-flux, in spatial and temporal directions. Both update steps are accepted with a 
Metropolis decision. The chemical potentials couple to temporally winding loops, giving larger weights to configurations 
with positive temporal winding numbers $\omega_\lambda [k]$. We remark that this update strategy of generating the conserved fluxes 
from plaquettes and defects (i.e., the winding loops) can be extended to a full Kramers-Wannier dualization
where the plaquettes and defect lines are introduced as the new variables \cite{Gattringer:2017hhn}. Furthermore we also experimented 
with a suitable worm strategy \cite{Giuliani:2017mxu}  for the update but observed severe inefficiencies for large $\beta$ and $\mu$. 
Possible improvements of the worm update strategy are currently part of our investigations.

\begin{figure}[t]
\begin{center}
\hspace*{20mm}
\includegraphics[scale=0.4,clip]{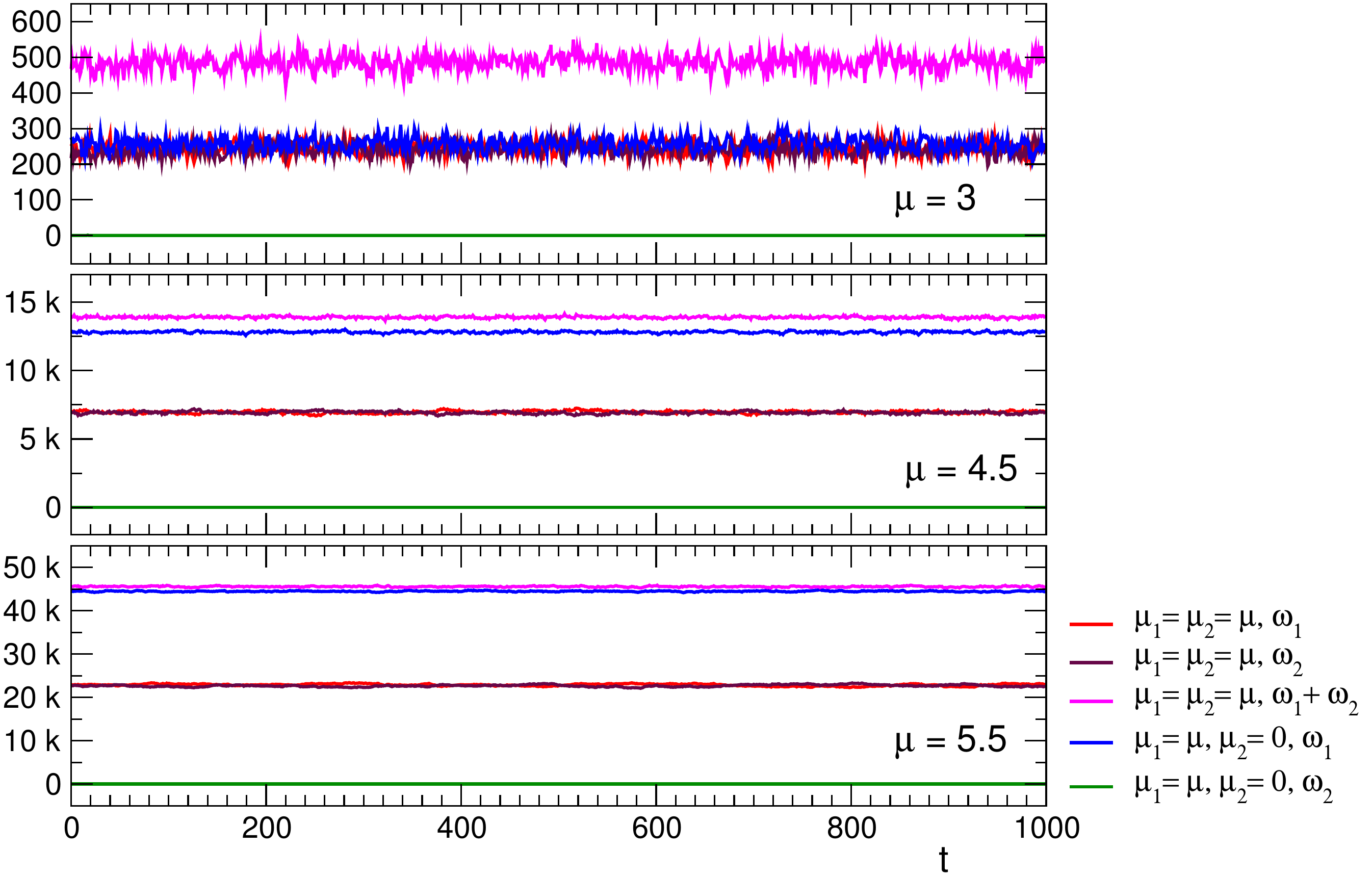}
\end{center}
\caption{Time series for the temporal winding numbers $\omega_\lambda$ (which correspond to the particle numbers) 
on $16^3 \times 4$ lattices at $J = 0.1$. We analyze two different scenarios:  $\mu_1 = \mu_2 = \mu$ and  
$\mu_1 = \mu$, $\mu_2 = 0$ for three different values of the parameter $\mu$.}
\label{fig:timeseries}	
\end{figure}

In Fig.~\ref{fig:chiUcompare} we show the results for the action density susceptibility $\chi_s$ as a function of the coupling $J$ at 
zero density on a $16^4$ lattice. The red circles represent the results obtained from a conventional simulation and the blue diamonds 
show the results of the worldline simulation performed with the local updates and confirm that the results from the two formulations 
agree well with each other. For the conventional simulation we used a statistics of $10^4$  measurements 
separated by two heat bath sweeps for decorrelation per coupling, after equilibrating with $10^3$ heat bath sweeps. 
In the worldline simulation we used for each coupling $10^4$ equilibration sweeps and a statistics of $2.5\times10^4$ measurements, 
separated by 5 decorrelation steps. 

After verifying the worldline simulation at $\mu_\lambda = 0$, we investigate the dynamics of the system at finite density. 
In Fig.~\ref{fig:timeseries} we show time series of the winding numbers as a function of the Monte Carlo time $t$, after an initial 
thermalization with $10^3$ update sweeps. We performed runs on a $16^3\times4$ ($N_t=4$) lattice at a fixed coupling of $J=0.1$ 
and for three different chemical potentials $\mu$, where we look at two different scenarios: $\mu_1 = \mu_2 = \mu$ and 
$\mu_1 = \mu$, $\mu_2 = 0$. Obviously the time series do not display long autocorrelations for either scenario.
The red and maroon lines represent the winding numbers $\omega_1$ and $\omega_2$ for the first scenario. 
As expected, they coincide for all three values of $\mu$ due to the equality of the chemical potentials. The blue and green lines 
represent the two winding numbers for the second scenario. Since in that case $\mu_2=0$, also the winding number of the second 
flavor $\omega_2$ vanishes, while $\omega_1$ has a non-vanishing value. The interesting question is whether the 
system favors a certain total winding number. In Fig.~\ref{fig:timeseries} we show the combination $\omega_1+\omega_2$ 
for the first scenario with a pink line. Comparing the plots, we see that for increasing $\mu$ the relative deviation of the 
total flux in the two scenarios decreases which indicates that the site couplings of the worldlines coming from the Haar measure
integration strongly ties together the two individual fluxes and generates a coherent dynamics of all worldline degrees of freedom
(compare the schematic illustration in Fig.~\ref{fig:worldlines}). 

\section{Dual formulation for SU(2) gauge theories}

Having presented the idea of dualization with abelian color fluxes for the SU(2) principal chiral model, the generalization of the
idea to SU(2) lattice gauge theory is more or less straightforward \cite{Gattringer:2016lml,Marchis:2016cpe}. We discuss the dualization 
in two steps, first for pure SU(2) lattice gauge theory and then for the case of SU(2) gauge fields coupled to staggered fermions.

\subsection{Dual formulation for pure SU(2) lattice gauge theory with abelian color cycles}

\begin{figure}[t]
\begin{center}
\vskip5mm
\includegraphics[scale=0.9,clip]{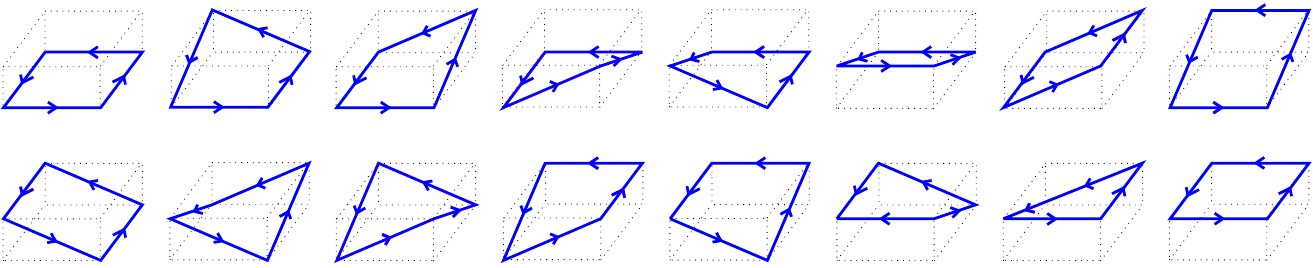}
\end{center}
\caption{The 16 abelian color cycles of SU(2): The ACCs are paths in color space closing around a plaquette. In our graphical 
representation the lower layer corresponds to color index 1, while the upper layer is color index 2. At each site of a plaquette 
the ACC can run through color 1 or 2 giving rise to the 16 different ACCs we show.}
\label{fig:allcycles}	
\end{figure}

The degrees of freedom of SU(2) lattice gauge theory are the 
link variables $U_{x,\mu} \in$ SU(2), assigned to the links of the lattice. The partition function is given by 
$Z = \int \! D[U] \, e^{-S_G[U]}$, where the path integral measure is the product of the invariant Haar measures 
$\int \! D[U] = \prod_{x,\mu} \int_{\text{SU}(2)} dU_{x,\mu}$ over all links. For the gauge 
action we use the Wilson form,
\begin{equation}
S_G[U] \; = \; -\dfrac{\beta}{2} \sum_{x,\mu < \nu} 
\mbox{Tr} \,  U_{x,\mu} \; U_{x+\hat{\mu},\nu} \, U_{x+\hat{\nu},\mu}^{\dagger} \, U_{x,\nu}^\dagger  \; = 
\; -\dfrac{\beta}{2} \sum_{x,\mu < \nu} \sum_{a,b,c,d=1}^{2} 
U_{x,\mu}^{ab} U_{x+\hat{\mu},\nu}^{bc} U_{x+\hat{\nu},\mu}^{dc \ \star} U_{x,\nu}^{ad \ \star} \; ,
\label{SU2_action}
\end{equation}
where in the second step we have already rewritten the trace and the matrix products as explicit sums over the SU(2) color 
indices. The partition function is then written as
\begin{eqnarray}
&& \hspace{20mm} Z  \; = \;  \int \! D[U] \prod_{x,\mu<\nu} \prod_{a,b,c,d = 1}^{2} 
e^{\frac{\beta}{2} U_{x,\mu}^{ab} U_{x+\hat{\mu},\nu}^{bc} U_{x+\hat{\nu},\mu}^{dc \ \star} U_{x,\nu}^{ad \ \star}} 
\nonumber \\
&& =  \int \! D[U] \prod_{x,\mu<\nu} \prod_{a,b,c,d = 1}^{2} \sum_{p_{x,\mu\nu}^{abcd} = 0}^{\infty} 
\dfrac{ \left( \beta/2 \right)^{p_{x,\mu\nu}^{abcd}}}{p_{x,\mu\nu}^{abcd}\, !} 
\left( U_{x,\mu}^{ab} U_{x+\hat{\mu},\nu}^{bc} U_{x+\hat{\nu},\mu}^{dc \ \star} 
U_{x,\nu}^{ad \ \star} \right)^{p_{x,\mu\nu}^{abcd}} \; ,
\label{eq:partitionsum}
\end{eqnarray}
where in the first step the exponential of the sum was rewritten as a product of individual exponentials. In the second step each of 
these individual exponentials was expanded in its Taylor series. The corresponding expansion indices  
$p_{x,\mu\nu}^{abcd} \in \mathds{N}_0$ appear as powers of the product 
$U_{x,\mu}^{ab} U_{x+\hat{\mu},\nu}^{bc} U_{x+\hat{\nu},\mu}^{dc \ \star} U_{x,\nu}^{ad \ \star}$ of elements of the link
variables. This product corresponds to a path in color space closing around a plaquette and is referred to as
abelian color cycle (ACC). The indices $p_{x,\mu\nu}^{abcd}$ we refer to as ''cycle occupation numbers''.
For SU(2) there are 16 different color cycles which are distinguished by the possible 
combinations of the color indices $a,b,c,d \in \{1,2\}$. In Fig.~\ref{fig:allcycles} we illustrate these 16 ACCs.

The ACCs $U_{x,\mu}^{ab} U_{x+\hat{\mu},\nu}^{bc} U_{x+\hat{\nu},\mu}^{dc \ \star} U_{x,\nu}^{ad \ \star}$ are complex numbers,
such that we can reorder them in (\ref{eq:partitionsum}) and organize them with respect to the individual links where they are 
integrated over with the Haar measure. Introducing the notation 
$\sum_{\{p\}} = \prod_{x,\mu<\nu} \prod_{a,b,c,d = 1}^{2} \sum_{p_{x,\mu\nu}^{abcd} = 0}^{\infty}$ for
the sum over all configurations of the cycle occupation numbers $p_{x,\mu\nu}^{abcd} \in \mathbb{N}_{0}$ and reordering the 
ACCs we find for the partition sum
\begin{equation}
Z \; = \; \sum_{\{p\}} \left[ \prod_{x,\mu<\nu} \prod_{a,b,c,d}  
\dfrac{ \left(\beta/2 \right)^{p_{x,\mu\nu}^{abcd}}}{p_{x,\mu\nu}^{abcd}\, !} \right] 
\prod_{x,\mu} \int \! \! dU_{x,\mu} \; \prod_{a,b} \left( U_{x,\mu}^{ab} \right) ^{N_{x,\mu}^{ab}} 
\left( U_{x,\mu}^{ab \ \star} \right) ^{\overline{N}_{x,\mu}^{ab}} \; ,
\label{eq:partitionsum2}
\end{equation}
where we collected the exponents $N_{x,\mu}^{ab}$ and 
$\overline{N}_{x,\mu}^{ab}$ for the matrix elements $U_{x,\mu}^{ab}$ and $U_{x,\mu}^{ab \ \star}$ given by
\begin{equation} 
N_{x,\mu}^{ab} \; = \; 
\sum_{\nu:\mu<\nu}p_{x,\mu\nu}^{abss} + \sum_{\rho:\mu>\rho}p_{x-\hat{\rho},\rho\mu}^{sabs}
\quad , \quad 
\overline{N}_{x,\mu}^{ab} \; = \; 
\sum_{\nu:\mu<\nu}p_{x-\hat{\nu},\mu\nu}^{ssba} + \sum_{\rho:\mu>\rho}p_{x,\rho\mu}^{assb} \; .
\end{equation} 
The labels $s$ indicate that the corresponding indices are summed independently, for example
$p_{x,\mu\nu}^{abss} \equiv \sum_{c,d} p_{x,\mu\nu}^{abcd}$ or 
$p_{x,\mu\nu}^{sabs} \equiv \sum_{c,d} p_{x,\mu\nu}^{cabd}$. In this form the integrals over the gauge link
matrix elements $U_{x,\mu}^{ab}$ and $U_{x,\mu}^{ab \ \star}$ can now be performed in closed form. Again we 
use the parameterization (\ref{pcm_parameterization}), but now the SU(2) matrices are assigned to the links. 
The partition function turns into
\begin{align}
&Z \; = \; \sum_{\{p\}} \! \left[ \prod_{x,\mu<\nu} \prod_{a,b,c,d} 
 \dfrac{ \left( \beta/2 \right)^{p_{x,\mu\nu}^{abcd}}}{p_{x,\mu\nu}^{abcd}!} \right]
 \prod_{x,\mu} (-1)^{J_{x,\mu}^{21}} \; \,
 2 \! \int_{0}^{\pi/2} \!\!\!\!  d\theta_{x,\mu} \, (\cos\theta_{x,\mu})^{1 + S_{x,\mu}^{11} + S_{x,\mu}^{22}} 
 \; (\sin\theta_{x,\mu})^{1 + S_{x,\mu}^{12} + S_{x,\mu}^{21}} 
 \nonumber
 \\
& \hspace{50mm} \times \; 
\int_{0}^{2\pi} \! \dfrac{d\alpha_{x,\mu}}{2\pi} \; e^{i\alpha_{x,\mu}[J_{x,\mu}^{11}-J_{x,\mu}^{22}]} \; 
\int_{0}^{2\pi} \! \dfrac{d\beta_{x,\mu}}{2\pi} \; e^{i\beta_{x,\mu}[J_{x,\mu}^{12}-J_{x,\mu}^{21}]} 	\; ,
\label{eq:partitionsum3}
\end{align} 
where we introduced the abbreviations $J_{x,\mu}^{ab} \; = \; N_{x,\mu}^{ab} - \overline{N}_{x,\mu}^{ab}$ and 
$S_{x,\mu}^{ab} \; = \; N_{x,\mu}^{ab} + \overline{N}_{x,\mu}^{ab}$.
The integrals over the phases $\alpha_{x,\mu}$ and $\beta_{x,\mu}$ give rise
to Kronecker deltas which enforce constraints for the fluxes $J_{x,\mu}^{ab}$ at all links $(x,\mu)$. These constraints are
\begin{equation}
J_{x,\mu}^{11} \, - \, J_{x,\mu}^{22} \; = \; 0 \quad \forall \, x, \mu \qquad \mbox{and} \qquad
J_{x,\mu}^{12} \, - \, J_{x,\mu}^{21} \; = \; 0 \quad \forall \, x, \mu \; .
\label{eq:constraints}
\end{equation}
As for the case of the SU(2) principal chiral model the constraints imply that the combinations 
$S_{x,\mu}^{11} \, + \, S_{x,\mu}^{22}$ and $S_{x,\mu}^{12} \, + \, S_{x,\mu}^{21}$ are even such that we can
again write the beta functions from the $\theta_{x,\mu}$-integrations as ratios of factorials. 

We obtain the final expression for the dual form of the partition function of SU(2) lattice gauge theory as a
sum over configurations of the plaquette occupation numbers $p_{x,\mu\nu}^{abcd} \in \mathbb{N}_{0}$ subject to constraints which
for a gauge theory are attached to the links of the lattice: 
\begin{equation}
Z \; = \; \sum_{\{p\}} W_{\beta}[p] \; W_H[p] \; (-1)^{\sum_{x,\mu}J_{x,\mu}^{21}} \; 
\prod_{x,\mu}  \delta(J_{x,\mu}^{11}-J_{x,\mu}^{22}) \; \delta(J_{x,\mu}^{12}-J_{x,\mu}^{21}) \; .
\label{SU2_finaldual}
\end{equation}
The configurations come with a sign factor that goes back to the parameterization of the SU(2) matrices. 
Whether it can be absorbed in a suitable resummation of terms in the dual representation is unclear at the moment. 
We also defined two weight factors  
\begin{equation}
\label{eq:weightfactorbeta}	
W_{\beta}[p] \; = \; 
\prod_{x,\mu<\nu} \prod_{a,b,c,d}  \dfrac{ \left( \frac{\beta}{2} \right)^{p_{x,\mu\nu}^{abcd}}}{p_{x,\mu\nu}^{abcd}!}  \quad , \quad	
W_{H}[p] \; = \; 
\prod_{x,\mu} \dfrac{\left(\frac{S_{x,\mu}^{11} + S_{x,\mu}^{22}}{2}\right)! 
\left(\frac{S_{x,\mu}^{12} + S_{x,\mu}^{21}}{2}\right)!}{\left(1 + \frac{S_{x,\mu}^{11} + S_{x,\mu}^{22} + 
S_{x,\mu}^{12} + S_{x,\mu}^{21}}{2}\right)!} \; .
\end{equation}
The $\beta$-dependent weight $W_{\beta}[p]$ collects the trivial factors from the 
expansion of the exponentials, while the weight factor $W_{H}[p]$ comes from the 
Haar measure integral and implements the SU(2) symmetry of the conventional representation in the dual form.

\begin{figure}[t]
\vspace{5mm}
\begin{center}
\hspace*{5mm}
\includegraphics[scale=0.7,clip]{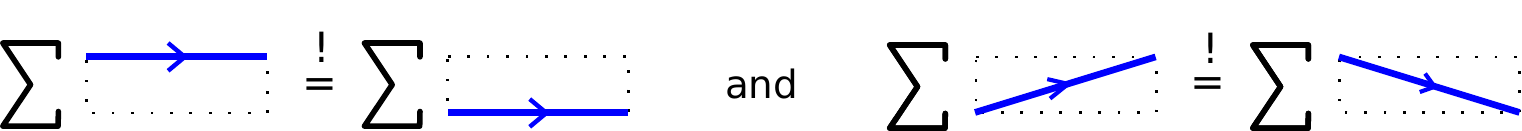}
\end{center}
\caption{Constraints for the fluxes $J_{x,\mu}^{ab}$: On all links the sum over all 1-1 flux 
must be equal to the sum over all 2-2 flux and the sum over 1-2 fluxes must equal the sum over 2-1 fluxes.
\hfill}
\label{fig:fluxconservation}
\end{figure}

To conclude this subsection, let us briefly discuss the structure of the constraints, which is interesting for comparison with the SU(3) case 
we address in the next section. The Kronecker deltas in (\ref{SU2_finaldual}) enforce
$J_{x,\mu}^{12} \, = \, J_{x,\mu}^{21}$ and $J_{x,\mu}^{11} \, = \, J_{x,\mu}^{22} \; \forall \, x, \mu$. The fluxes 
$J_{x,\mu}^{ab}$ collect the contributions of all Abelian color cycles that connect color $a$ at $x$ with color $b$ at $x + \hat{\mu}$. 
Thus the constraint  $J_{x,\mu}^{12} \, = \, J_{x,\mu}^{21}$ enforces that along the link $x, \mu$ the flux from color 1 to color 2 is
the same as the total flux from color $2$ to color $1$. The constraint $J_{x,\mu}^{11} \, = \, J_{x,\mu}^{22}$ enforces that along that link 
the flux between equal colors has to be the same. Thus together the two constraints imply that for both colors the flux is the same for
both links, and that along links flux can be exchanged between the two colors. We will find a more general but similar set of constraints
for the case of SU(3).

\subsection{Coupling staggered fermions}

Let us briefly discuss the coupling to matter fields and how they can be treated with Abelian color fluxes. More specifically we consider
the partition function $Z = \int \! D[U] \, e^{-S_G[U]} Z_F[U]$ where $Z_F[U] = \int \! D[\psi,\overline{\psi}] \, e^{-S_F[\psi,\overline{\psi},U]}$
is the partition function of staggered fermions in a background of SU(2) gauge fields. The measure $\int \! D[\psi,\overline{\psi}]$ is the 
product measure over Grassmann variables $\psi_x^a$ and $\overline{\psi}_x^{\,a}$ attached to the sites $x$. 
The corresponding fermion action is
\begin{equation}
S_F[\,\overline{\psi}, \psi,U]  \; = \;
\sum_x \Big[ \, m \!\sum_{a} 
\overline{\psi}_x^{\,a} \psi_x^{a} + \sum_{\mu} \frac{\gamma_{x,\mu}}{2} \sum_{a,b}
\Big( \, \overline{\psi}^{\,a}_x U_{x,\mu}^{\,ab} 
\psi_{x + \hat\mu}^{b} - \overline{\psi}_{x+\hat\mu}^{\,b} U_{x,\mu}^{\, ab \, \star} \psi_{x}^{a} 
\Big) \Big] \; ,
\label{fermionaction}
\end{equation}
where $\gamma_{x,\mu}$ denotes the staggered sign factor and we have written all sums over color indices explicitly.
Again we turn all sums in the exponent into products and expand each individual Boltzmann factor: 
\begin{eqnarray}
\hspace*{-6mm} && Z_F[U] \, = \int \!\!\! D[\, \overline{\psi}, \psi]  \prod_x \prod_a e^{-m \overline{\psi}_x^{\,a} \psi_x^a} \; 
\prod_{x,\mu} \prod_{a,b} e^{- \,\frac{\gamma_{x,\mu}}{2} \overline{\psi}^{\,a}_x U_{x,\mu}^{\,ab} \psi_{x + \hat\mu}^b} \;
e^{ \, \frac{\gamma_{x,\mu}}{2} \overline{\psi}_{x+\hat\mu}^{\,b} U_{x,\mu}^{\, ab\, \star} \psi_{x}^a} 
\nonumber \\ 
\hspace*{-6mm} && = \int \!\!\!D[\, \overline{\psi}, \psi]  \prod_x \prod_a \sum_{s_x^a = \, 0}^{1} \! 
( -m \overline{\psi}_x^{\,a} \psi_x^a )^{s_x^a} 
\prod_{x,\mu} \prod_{a,b} 
\sum_{k_{x,\mu}^{\;ab} = \, 0 }^{1} \!\!\! 
(- \frac{\gamma_{x,\mu}}{2} \, \overline{\psi}^{\,a}_x U_{x,\mu}^{\,ab} \psi_{x + \hat\mu}^b)^{k_{x,\mu}^{\; ab}} \!
\sum_{ \overline{k}_{x,\mu}^{\;ab} = \, 0}^{1}\!\!\! 
( \frac{\gamma_{x,\mu}}{2} \, \overline{\psi}_{x+\hat\mu}^{\,b} U_{x,\mu}^{\, ab\, \star} 
\psi_{x}^a)^{\overline{k}_{x,\mu}^{\; ab}}
\nonumber \\
\hspace*{-6mm} &&= \, \frac{1}{2^{2V}} \! \sum_{\{s,k,\overline{k}\}} \!\! (2m)^{\, \sum_{x,a} s_x^a} \; \prod_{x,\mu} \prod_{a,b} 
( U_{x,\mu}^{\,ab} )^{k_{x,\mu}^{\; ab}} \, 
( U_{x,\mu}^{\, ab\, \star} )^{\overline{k}_{x,\mu}^{\; ab}}  \; (-1)^{k_{x,\mu}^{\; ab}} \; 
( \, \gamma_{x,\mu}\,) ^{k_{x,\mu}^{\; ab} + \overline{k}_{x,\mu}^{\; ab} } 
\nonumber \\
\hspace*{-6mm} && \hspace{40mm} \times \!\!
\int \!\! D[\, \overline{\psi}, \psi] \! \prod_x \prod_a  (\overline{\psi}_x^{\,a} \psi_x^a )^{s_x^a} \prod_{x,\mu} \prod_{a,b}
(\overline{\psi}^{\,a}_x \psi_{x + \hat\mu}^b)^{k_{x,\mu}^{\; ab}} 
(\overline{\psi}_{x+\hat\mu}^{\,b} \psi_{x}^a)^{\overline{k}_{x,\mu}^{\; ab}} \; .
\label{fermiondual1}
\end{eqnarray}
We have introduced summation variables $k_{x,\mu}^{\; ab}$ for the hopping terms that connect color $a$ at $x$ with color $b$ at 
$x + \hat{\mu}$ and $\overline{k}_{x,\mu}^{\; ab}$ for the corresponding inverse hop, as well as variables $s_x^a$ for the contribution
of color $a$ to the mass term. Due to the nilpotency of the Grassmann numbers all these variables can 
only have the values $0$ and $1$ and may 
be viewed as activation numbers for the corresponding terms. 

After some reordering of terms we have isolated the remaining Grassmann integral in the last line of (\ref{fermiondual1}). This integral 
depends on the values of the activation numbers and can be $0,-1$ or $+1$. A non-zero value is possible only for certain configurations 
of the activation numbers: The forward and backward hopping terms can form closed loops in both, the space-time lattice and in color 
space, similar to the Abelian color cycles depicted in Fig.~\ref{fig:allcycles}. Of course not only plaquettes are possible, 
but also longer loops, as well as so-called dimers where a hop and its inverse are combined. 
All loops and dimers come with the corresponding 
entries $U_{x,\mu}^{\,ab}$ and $U_{x,\mu}^{\,ab \, \star}$ of the link variables along their contour and we refer to this structure as
the abelian color fluxes already mentioned before. Finally also with the activation numbers $s_x^a$ for the mass terms we can
saturate the Grassmann integral, and we refer to those terms as monomers. 

We can summarize the constraint for obtaining a non-zero result for the Grassmann integral as follows: Again we consider the lattice as 
a product of two space-time lattices, with one lattice for each of the factors. The Grassmann integral is non-zero if each site of that double 
lattice is either occupied by a monomer, is run through by a loop or is the endpoint of a dimer. For staggered fermions one can also work 
out the sign for these non-trivial configurations. We do not address the sign determination here and refer to \cite{Gattringer:2016lml} for
this step. The factors with the gauge link elements $U_{x,\mu}^{\,ab}$ and $U_{x,\mu}^{\,ab \, \star}$ now have to be combined with
the corresponding factors in (\ref{eq:partitionsum2}) and are then integrated with $\int \! D[U]$. Thus the constraints and the weights now 
also depend on the activation numbers of the fermions.  We find the following result for the dual representation of the partition function:
\begin{eqnarray}
Z & = &\frac{1}{2^{2V}} \!\!\sum_{\{p,k,\overline{k},s \}} \!\! C_{F}[s,k,\overline{k}\,] \;  W_M[s] \, W_\beta [p] \, 
W_H [p,k,\overline{k}]\; 
\prod_{x,\mu} (-1)^{\, J_{x,\mu}^{\,21} + k_{x,\mu}^{\,21} + \overline{k}_{x,\mu}^{\,21}} \; 
\prod_{L} \, \mbox{sign} \,(L) 
\nonumber \\
& \times & \!\!\!\!\!
\prod_{x, \mu} 
\delta \left( J_{x,\mu}^{\,11} \!+\! k_{x,\mu}^{\,11} \!-\! \overline{k}_{x,\mu}^{\,11} - 
[J_{x,\mu}^{\,22} \!+\! k_{x,\mu}^{\,22} \!-\! \overline{k}_{x,\mu}^{\,22} ]  \right)
\delta \left( J_{x,\mu}^{\,12} \!+\! k_{x,\mu}^{\,12} \!-\! \overline{k}_{x,\mu}^{\,12} - 
[J_{x,\mu}^{\,21} \!+\! k_{x,\mu}^{\,21} \!-\! \overline{k}_{x,\mu}^{\,21} ]  \right).
\label{dualZfinal}
\end{eqnarray}
The partition function now is a sum over all configurations of the fermion activation numbers $k_{x,\mu}^{\; ab}$, 
$\overline{k}_{x,\mu}^{\; ab}$ and $s_x^a$, as well as the cycle occupation numbers $p_{x,\mu\nu}^{abcd}$. 
The fermion activation numbers come with a fermion constraint function $C_{F}[s,k,\overline{k}\,]$ which equals 1 for all configurations
where each site is either run through by a loop, is the endpoint of a dimer or is occupied by a monomer. Otherwise 
$C_{F}[s,k,\overline{k}\,] = 0$. The monomer weight function $W_M[s]$ collects all the mass term contributions and has the simple form
$W_M[s] = (2m)^{\, \sum_{x,a} s_x^a}$. The weight factor $W_\beta [p]$ remains as in the case of pure gauge theory 
(Eq.~\ref{eq:weightfactorbeta}), while the weight factor $W_H [p,k,\overline{k}]$ 
from the Haar measure integration now also collects the contributions from the link terms in the fermion action. It still has the
form (\ref{eq:weightfactorbeta}), but the arguments $S_{x,\mu}^{ab}$ now also depend on the activation of 
fermion hops, i.e., $S_{x,\mu}^{ab} \; = \; N_{x,\mu}^{ab} + \overline{N}_{x,\mu}^{ab} + k_{x,\mu}^{ab} + \overline{k}_{x,\mu}^{ab}$.

In addition to the signs from the 2-1 components in the SU(2) parameterization (\ref{pcm_parameterization}), 
also the Grassmann integral contributes to the overall sign. This sign depends only on the loops $L$ of the dual fermions variables
and for a loop $L$ is given by sign$ (L) \; = \; (-1)^{ \,1 \, + \, |L|/2 \, + \, W_L \, + \,  P_L}$. 
Here $|L|$ is the length of the loop $L$, $W_L$ its temporal winding number, and
$P_L$ the total number of plaquettes needed to fill the loop with a surface such that $L$ appears as the boundary 
of the surfaces (see \cite{Gattringer:2016lml} for a detailed discussion of the sign factor). Finally, the constraints take into account
also the Abelian color fluxes from the fermion hopping terms, as can be seen by the corresponding activation numbers $k_{x,\mu}^{\; ab}$, 
$\overline{k}_{x,\mu}^{\; ab}$ in the Kronecker deltas. The constraints are still as illustrated in Fig.~\ref{fig:fluxconservation},
but the sums are now over the contributions of all ACCs attached to a link, as well as all fermion fluxes.

The form of the ACC/ACF representation of the partition sum in (\ref{dualZfinal}) is very satisfactory from a conceptual point of
view. SU(2) lattice gauge theory is exactly represented as a sum over cycle and loop configurations that correspond to
closed paths in space time as well as in color. The symmetries of the conventional form appear as constraints for the dual variables
and local interaction terms. We remark that one can also couple a chemical potential to the fermions, which as in the case of the
principal chiral model would couple to the temporal winding number of the fermion loops, underlining again the highly geometrical
nature of the dual representation. However, for a numerical simulation a suitable resummation of negative sign contributions
has to be found, as we already remarked in the previous subsection for the case of pure SU(2) lattice gauge theory. In this 
context it is interesting to note that the leading terms in the strong coupling series all have positive signs \cite{Gattringer:2016lml}
indicating that such a resummation strategy might exist.

\section{Dual formulation for SU(3) gauge theories}

We conclude this proceedings contribution with a discussion of the challenges and new features that appear when one switches to the gauge group SU(3). Also in this case we start with the derivation of the worldsheet representation for pure SU(3) lattice gauge theory, and then discuss the ACC/ACF dual representation for full QCD.

\subsection{Dual formulation for pure SU(3) lattice gauge theory with abelian color cycles} 

The generalization to the SU(3) gauge group of the abelian color cycle (ACC) dualization procedure discussed in Sec.~3.1 starts by noticing that, because of the non pseudo-reality of SU(3), in the Wilson action we have to take the real part of the trace of the plaquette variables
explicitly: 
\begin{align}
\label{eq:actionsu3}
S_G[U] \; &= \; -\dfrac{\beta}{3} \sum_{x,\mu < \nu} \!
\textnormal{Re} \textnormal{Tr} \left[ \,U_{x,\mu} \; U_{x+\hat{\mu},\nu}  U_{x+\hat{\nu},\mu}^{\dagger}  U_{x,\nu}^\dagger \, \right] \; =\; -\dfrac{\beta}{6} \sum_{x,\mu < \nu}  \sum_{a,b,c,d=1}^{3} \left[ \,
U_{x,\mu}^{ab} U_{x+\hat{\mu},\nu}^{bc} U_{x+\hat{\nu},\mu}^{dc \, \star} U_{x,\nu}^{ad \, \star} \, + \,c.c. \,\right] .
\end{align}
This results in ACCs that can either have mathematically positive 
($U_{x,\mu}^{ab} U_{x+\hat{\mu},\nu}^{bc} U_{x+\hat{\nu},\mu}^{dc \, \star} U_{x,\nu}^{ad \, \star}$) 
or negative ($U_{x,\mu}^{ab \, \star} U_{x+\hat{\mu},\nu}^{bc \, \star} U_{x+\hat{\nu},\mu}^{dc} U_{x,\nu}^{ad}$) 
orientation. Both types of ACCs are again complex numbers that may be viewed as cycles along 
the plaquettes $(x,\mu\nu)$ of the lattice with a path in color space labelled by the indices $a,b,c,d$. 
For SU(3) each of these indices has 3 possible values such that 
there is a total of 81 different ACCs. Another consequence of the non pseudo-reality of SU(3) is that 
when we Taylor expand the locally factorized Boltzmann weight we need two sets of summation variables, 
$n_{x,\mu\nu}^{abcd} \in \mathbb{N}_{0}$ and $\overline{n}_{x,\mu\nu}^{abcd} \in \mathbb{N}_{0}$, that 
correspond to positive and negative oriented ACCs respectively:
\begin{align}
\label{eq:partitionsumsu3}
Z \; &= \int \! D[U] \prod_{x,\mu<\nu} \prod_{a,b,c,d = 1}^{3} 
e^{\,\frac{\beta}{6} \, ( U_{x,\mu}^{ab} U_{x+\hat{\mu},\nu}^{bc} U_{x+\hat{\nu},\mu}^{dc \, \star} U_{x,\nu}^{ad \, \star} )} \;
e^{\, \frac{\beta}{6} \, ( U_{x,\mu}^{ab \, \star} U_{x+\hat{\mu},\nu}^{bc \, \star} U_{x+\hat{\nu},\mu}^{dc} U_{x,\nu}^{ad} )}
\\
\nonumber
& = \int \! D[U] \prod_{x,\mu<\nu} \prod_{a,b,c,d = 1}^{3} \sum_{n_{x,\mu\nu}^{abcd} = 0}^{\infty} \sum_{\bar{n}_{x,\mu\nu}^{abcd} = 0}^{\infty} 
\dfrac{ \left( \beta/6 \right)^{n_{x,\mu\nu}^{abcd} + \bar{n}_{x,\mu\nu}^{abcd}}}{n_{x,\mu\nu}^{abcd}! \; \bar{n}_{x,\mu\nu}^{abcd}\, !} 
\left( U_{x,\mu}^{ab} U_{x+\hat{\mu},\nu}^{bc} U_{x+\hat{\nu},\mu}^{dc \, \star} 
U_{x,\nu}^{ad \, \star} \right)^{n_{x,\mu\nu}^{abcd}}
\Big( c. c. \Big)^{\bar{n}_{x,\mu\nu}^{abcd}} \\
\nonumber
&= \; \sum_{\{n, \bar{n}\}} \left[ \prod_{x,\mu<\nu} \prod_{a,b,c,d}  
\dfrac{ \left( \beta/6 \right)^{n_{x,\mu\nu}^{abcd} + \bar{n}_{x,\mu\nu}^{abcd}}}{n_{x,\mu\nu}^{abcd}! \; \bar{n}_{x,\mu\nu}^{abcd}\, !} \right]  \prod_{x,\mu} \int \! \! dU_{x,\mu} \; \prod_{a,b} \left( U_{x,\mu}^{ab} \right) ^{N_{x,\mu}^{ab}} 
\left( U_{x,\mu}^{ab \ \star} \right) ^{\overline{N}_{x,\mu}^{ab}} \; .
\end{align}
The steps in (\ref{eq:partitionsumsu3}) are analogous to the ones performed in (\ref{eq:partitionsum}) and in (\ref{eq:partitionsum2}), but here the integer valued powers $N_{x,\mu\nu}^{abcd}$ and $\bar{N}_{x,\mu\nu}^{abcd}$ for the matrix elements $U_{x,\mu}^{ab}$ and $U_{x,\mu}^{ab \ \star}$ in the last line are
\begin{gather} 
\label{eq:N}
N_{x,\mu}^{ab} = \!
\sum_{\nu:\mu<\nu} \left[n_{x,\mu\nu}^{abss} + \bar{n}_{x-\hat{\nu},\mu\nu}^{ssba}\right] + \! \!
\sum_{\rho:\mu>\rho}\left[\bar{n}_{x,\rho\mu}^{assb} + n_{x-\hat{\rho},\rho\mu}^{sabs}\right] , \ \
\overline{N}_{x,\mu}^{ab}  = \!
\sum_{\nu:\mu<\nu} \left[\bar{n}_{x,\mu\nu}^{abss} + n_{x-\hat{\nu},\mu\nu}^{ssba}\right]  +  \! \!
\sum_{\rho:\mu>\rho} \left[n_{x,\rho\mu}^{assb} + \bar{n}_{x-\hat{\rho},\rho\mu}^{sabs}\right] ,
\end{gather} 
where again the label $s$ stands for the independent sum over the color indices that are replaced by $s$.

The final form of the partition sum (\ref{eq:partitionsumsu3}) is obtained by choosing an explicit parametrization for the SU(3) matrices \cite{Bronzan:1988wa},
\begin{gather}
\label{eq:parametrization}
U_{x,\mu} =\left(
\begin{array}{ccc}
c_1 c_2 \, e^{i\phi_1}  & s_1 \, e^{i\phi_3} & c_1 s_2 \, e^{i\phi_4}\\
s_2 s_3 \, e^{-i\phi_4 -i\phi_5} - s_1 c_2 c_3 \, e^{i\phi_1 +i\phi_2 -i\phi_3} & c_1 c_3 \, e^{i\phi_2} & - c_2 s_3 \, e^{-i\phi_1 -i\phi_5} - s_1 s_2 c_3 \, e^{i\phi_2 -i\phi_3 +i\phi_4}\\
- s_2 c_3 \, e^{-i\phi_2 -i\phi_4} - s_1 c_2 s_3 \, e^{i\phi_1 -i\phi_3 +i\phi_5} & c_1 s_3 \, e^{i\phi_5} & c_2 c_3 \, e^{-i\phi_1 -i\phi_2} - s_1 s_2 s_3 \, e^{-i\phi_3 +i\phi_4 +i\phi_5}
\end{array} \right) \; ,
\end{gather}
where $c_{i} = \cos \theta_{x,\mu}^{(i)}$, $s_{i} = \sin \theta_{x,\mu}^{(i)}$, with 
$\theta_{x,\mu}^{(i)} \in [0, \pi/2]$, and $\phi_{i} = \phi_{x,\mu}^{(i)}$, with $\phi_{x,\mu}^{(i)} \in [-\pi,\pi]$. The normalized Haar measure is 
$dU_{x,\mu}  =  1/(2\pi^5) \, d\theta_1 c_1^3 s_1 \, d\theta_2 c_2 s_2 \, d\theta_3 c_3 s_3 \, d\phi_1 \, d\phi_2 \, d\phi_3 \, d\phi_4 \, d\phi_5$ .
The next step is to insert this expression into the product measure $D[U] = \prod_{x,\nu} dU_{x,\nu}$, 
and the matrix elements $U_{x,\nu}^{ab}$ into (\ref{eq:partitionsumsu3}). For those matrix elements that are expressed as sums of 
complex numbers we use the binomial theorem $(x + y)^{N} = \sum_{m = 0}^{N} \binom{N}{m} \,x^{\,N-m}\, y^{\,m}$, thus introducing 
new auxiliary variables $m_{x,\nu}^{ab} \in \{0, N_{x,\nu}^{ab}\}$ for the matrix elements $(a,b) \in \{(2,1),(2,3),(3,1),(3,3)\}$ and 
$\overline{m}_{x,\nu}^{ab} \in \{0, \overline{N}_{x,\nu}^{ab}\}$ for the respective complex conjugate. 
It is convenient to perform a change of variables,
\begin{gather}
\label{eq:cycle}
n_{x,\mu\nu}^{abcd} \,-\, \bar{n}_{x,\mu\nu}^{abcd} \,=\, p_{x,\mu\nu}^{abcd} \ , \quad p_{x,\mu\nu}^{abcd} \in \mathbb{Z} \quad ; \qquad 
n_{x,\mu\nu}^{abcd} \,+\, \bar{n}_{x,\mu\nu}^{abcd} \,=\, |p_{x,\mu\nu}^{abcd}| \,+\, 2 l_{x,\mu\nu}^{abcd} \ , \quad l_{x,\mu\nu}^{abcd} \in \mathbb{N}_0 \ ,
\end{gather}
and to introduce the fluxes
\begin{gather}
\label{eq:Jfluxes}
J_{x,\mu}^{ab} \; =  
\sum_{\nu:\mu<\nu}  \left[ p_{x,\mu\nu}^{abss} - p_{x-\hat{\nu},\mu\nu}^{ssba} \right] - \!\!
\sum_{\rho:\mu>\rho} \left[ p_{x,\rho\mu}^{assb} - p_{x-\hat{\rho},\rho\mu}^{sabs} \right] ,
\\
\label{eq:Sfluxes}
S_{x,\mu}^{ab} =  
\sum_{\nu:\mu<\nu}  \left[ |p_{x,\mu\nu}^{abss}| + |p_{x-\hat{\nu},\mu\nu}^{ssba}| + 
2(l_{x,\mu\nu}^{abss} + l_{x-\hat{\nu},\mu\nu}^{ssba}) \right] 
\, + \sum_{\rho:\mu>\rho} \left[ |p_{x,\rho\mu}^{assb}| + |p_{x-\hat{\rho},\rho\mu}^{sabs}| + 
2(l_{x,\rho\mu}^{assb} + l_{x-\hat{\rho},\rho\mu}^{sabs}) \right] .
\end{gather}

\begin{figure}[t]	
	\centering	
		\includegraphics[scale=0.7,clip]{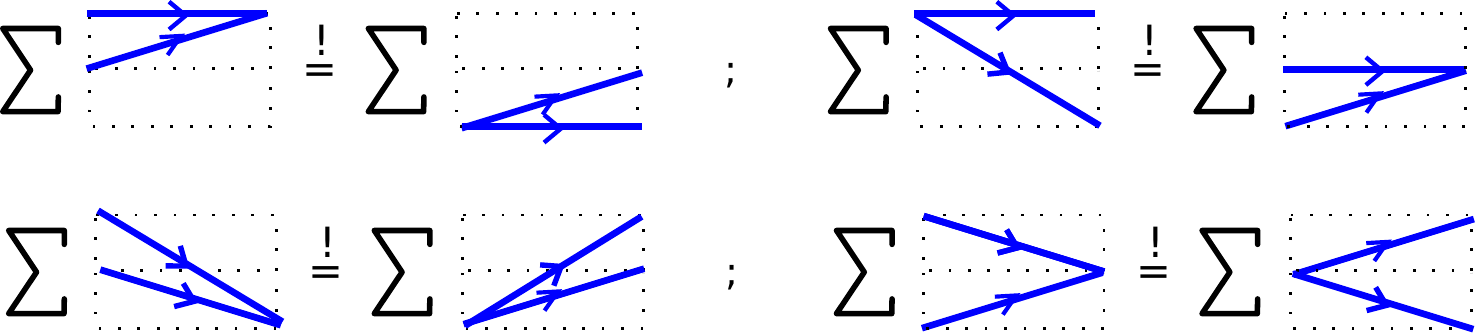}
	\caption{Graphical representation of the constraints for the fluxes in the SU(3) case. The constraints depicted in the first row are a 
	generalization of those for the SU(2) case (compare Fig.~5), as they relate fluxes between different colors. The constraints in 
	the second row enforce flux conservation, since they require that for each link the flux out of a color has to match the flux 
	into that color.
				\hfill}
	\label{fig:fluxconservation_SU3}
\end{figure}

At this point the Haar measure integrals can be solved in closed form. In particular, the integrations over the $\theta_{i}$ angles give rise to 
combinatorial factors, which we collect in the weight factor $W_{H}[p,l,m,\overline{m}]$, whose complete expression will be presented
in a forthcoming publication, while integrating the phases $\phi_{i}$ gives rise to constraints $C_{H}[p,m,\overline{m}]$ 
expressed in terms of Kronecker deltas,
\begin{align}
\label{eq:CH}
C_{H} [p,m,\overline{m}] \;& = \; \prod_{x,\mu} \delta(J_{x,\mu}^{11} + J_{x,\mu}^{12} - J_{x,\mu}^{33} - J_{x,\mu}^{23}) \;
\delta(J_{x,\mu}^{22} + J_{x,\mu}^{12} - J_{x,\mu}^{33} - J_{x,\mu}^{31}) \; \\
\nonumber 
& \times  \; \delta(J_{x,\mu}^{12} - j_{x,\mu}^{21} - j_{x,\mu}^{23} - j_{x,\mu}^{31} - j_{x,\mu}^{33}) \; 	
\delta(J_{x,\mu}^{13} + J_{x,\mu}^{12} - J_{x,\mu}^{31} - J_{x,\mu}^{21}) \; 
\delta(J_{x,\mu}^{32} + J_{x,\mu}^{12} - J_{x,\mu}^{23} - J_{x,\mu}^{21}) \; .
\end{align}
We introduced $j_{x,\nu}^{\,ab} = m_{x,\nu}^{ab} - \overline{m}_{x,\nu}^{ab}$ to simplify the notation. These constraints are graphically 
represented in Fig. \ref{fig:fluxconservation_SU3}. They relate the auxiliary variables $m$ and $\overline{m}$ to the cycle occupation 
numbers $p$ and enforce relations between combinations of the cycle occupation numbers. They are a generalized form of 
the constraints  
for SU(2) to a lattice with three layers, and again their effect is to restrict the admissible configurations to those that fulfill the flux 
conservation laws implied by (\ref{eq:CH}). 

The final form of the partition function,
\begin{equation}
\label{eq:partitionsumsu32}
Z \; = \; \sum_{\{p,l\}} \sum_{\{m,\overline{m}\}} W_{\beta}[p,l] \; W_H[p,l,m,\overline{m}] \; C_{H} [p,m,\overline{m}] \;
(-1)^{\, \sum_{x,\mu}J_{x,\mu}^{12} + J_{x,\mu}^{23} + J_{x,\mu}^{31} - j_{x,\mu}^{\,23} - j_{x,\mu}^{\,31}}    \; ,
\end{equation}
is a sum over the configurations of the new dual variables: the cycle occupation numbers $p_{x,\mu\nu}^{abcd} \in \mathbb{Z}$, and the 
auxiliary variables $l_{x,\mu\nu}^{\,abcd} \in \mathbb{N}_{0}$ and 
$m^{ab}_{x,\mu} \in \{0, N^{ab}_{x,\mu}\}$, $\overline{m}^{ab}_{x,\mu} \in \{0, \overline{N}^{ab}_{x,\mu}\}$, 
with $(a,b) \in \{(2,1),(2,3),(3,1),(3,3)\}$. Admissible configurations have to satisfy the link flux constraints 
(\ref{eq:CH}). Each configuration comes with the positive weights $W_{\beta}[p,l]$ and $W_{H}[p,l, m, \overline{m}]$, and with sign 
factors $(-1)^{\, \sum_{x,\mu}J_{x,\mu}^{12} + J_{x,\mu}^{23} + J_{x,\mu}^{31} - j_{x,\mu}^{\,23} - j_{x,\mu}^{\,31}}$, which origin 
from the SU(3) parametrization. 

\subsection{Dual representation of full lattice QCD}
\label{QCD}

The coupling of matter fields proceeds in the same way as we discussed for SU(2). We again use 
the staggered discretization of the fermion 
action (\ref{fermionaction}), and the dualization of the fermion partition sum reads as in (\ref{fermiondual1}), with the only difference that for 
SU(3) the color labels run from 1 to 3. The power of the overall factor in front of the sum over the configurations of the fermion dual 
variables $\{s,k, \overline{k}\}$ here is $3V$ instead of $2V$. 
The interpretation of the fermion's dual variables is analogous to the one we gave in Sec.~3.2: $s_{x}^{a} = 0, 1$ corresponds to the color 
component $a$ of the mass term on site $x$. $k_{x,\nu}^{ab} = 0, 1$ represents the forward hop from color $a$ to color $b$ on the link 
$(x,\nu)$, and $\overline{k}_{x,\nu}^{ab} = 0, 1$ is the respective backward hop on the same link. 

The Grassmann integral in the last line of (\ref{fermiondual1}) is non-vanishing only if each Grassmann variable 
$\psi_{x}^{a} \overline{\psi}_{x}^{a}$ appears exactly once and we express this condition with the fermion constraint 
$C_{F}[s,k,\overline{k}]$. 
As discussed in Sec.~3.2, this constraint can be fulfilled by occupying all the three layers of every site of our lattice with monomers, 
dimers, or loops. These are the only admissible fermion configurations.  

If the constraint $C_{F}[s,k,\overline{k}]$ is satisfied, the Grassmann integral in Eq.~(\ref{fermiondual1}) gives $+ 1$ or $-1$. 
Other signs are generated by the staggered sign factors 
$\prod_{x,\nu} (\eta_{x,\nu})^{\sum_{a,b} [k_{x,\nu}^{ab} + \overline{k}_{x,\nu}^{ab}] }$, the activation of forward hops $\prod_{x,\nu} 
(-1)^{\sum_{a,b} k_{x,\nu}^{ab}}$, and the anti-periodic boundary conditions along the time direction. They are the same as for 
the SU(2) case and one finds that only loop configurations introduce signs. 
We can express the sign contribution of every loop $L$ in the same simple form we found for 
SU(2), 
$
\textnormal{sign}(L) = (-1)^{1 + |L|/2 + P_L + W_L}
$
where $|L|$ is the length of the loop $L$, $P_L$ is the number of plaquettes necessary to cover the surface bounded by the loop $L$, 
and $W_L$ is the number of temporal windings. 

After the Grassmann integration the partition sum for QCD reads
\begin{equation}
\label{eq:partition}
Z \;=\; \dfrac{1}{2^{3V}}\sum_{\{s,k,\overline{k}\}} C_{F}[s,k,\overline{k}] \; W_{M}[s] \; W_{\mu}[k,\overline{k}] \;
\prod_{L} \textnormal{sign}(L) \int \! D[U] \, e^{S_{G}[U]}
\prod_{x, \nu} \prod_{a, b} 	\left( U_{x,\nu}^{ab} \right)^{k_{x,\nu}^{ab}}
\left( U_{x,\nu}^{ab \, \star} \right)^{\overline{k}_{x,\nu}^{ab}} \, ,
\end{equation}
where 
$
W_{M}[s] = (2m)^{\, \sum_{x,a} s_{x}^{a}}$ is the monomer weight factor and 
$W_{\mu}[k,\overline{k}] = e^{\, \mu \sum_{x}\sum_{ab}[k_{x,\hat{4}}^{ab} - \overline{k}_{x,\hat{4}}^{ab} ]}$
is the $\mu$-dependent weight factor. To obtain the final form of $Z$ we only have to perform the Haar integration 
in (\ref{eq:partition}). This differs from the one discussed in the last section just by the presence of the factors
$
\prod_{x, \nu} \prod_{a, b} 	\left( U_{x,\nu}^{ab} \right)^{k_{x,\nu}^{ab}}
\left( U_{x,\nu}^{ab \, \star} \right)^{\overline{k}_{x,\nu}^{ab}} \, ,
$
whose effect modifies the expression of the combinatorial factors collected in the weight $W_{H}[p,l,k,\overline{k},m,\overline{m}]$ and of 
the constraints $C_{H}[p,k,\overline{k},m,\overline{m}]$, but not the procedure to obtain them. In particular, the new weights and the 
new constraints can be obtained from the ones in the pure gauge case with the substitutions 
$J_{x,\mu}^{ab} \rightarrow J_{x,\mu}^{ab} + k_{x,\nu}^{ab} - \overline{k}_{x,\nu}^{ab}$ and 
$S_{x,\mu}^{ab} \rightarrow S_{x,\mu}^{ab} + k_{x,\nu}^{ab} + \overline{k}_{x,\nu}^{ab}$.

Thus for the dual partition sum of full QCD with staggered fermions we obtain
\begin{align}
\nonumber
Z \; = \; \frac{1}{2^{3V}} \sum_{\{p,l\}} \sum_{\{s,k,\overline{k}\}} \sum_{\{m,\overline{m}\}} & W_{\beta}[p,l] \; W_H[p,l,k,\overline{k},m,\overline{m}] \;W_{M}[s] \; W_{\mu}[k,\overline{k}] \;  C_{F}[s,k,\overline{k}] \; C_{H}[p,k,\overline{k},m,\overline{m}] \;  \\
\label{eq:partitionQCD2}
& \times \prod_{L} \textnormal{sign} (L) \prod_{x,\nu} (-1)^{J_{x,\nu}^{12} + K_{x,\nu}^{12} + J_{x,\nu}^{23} + K_{x,\nu}^{23}  + 
J_{x,\nu}^{31} + K_{x,\nu}^{31}  - j_{x,\nu}^{\,23} - j_{x,\nu}^{\,31}}   \; ,
\end{align}
where we use the short hand notation $K_{x,\nu}^{ab} = k_{x,\nu}^{ab} - \overline{k}_{x,\nu}^{ab}$. 
Eq.~(\ref{eq:partitionQCD2}) is an exact rewriting of the partition function of QCD with staggered fermions in terms of integer 
valued dual variables: The cycle occupation numbers $p_{x,\mu\nu}^{abcd} \in \mathbb{Z}$, and the auxiliary variables 
$l_{x,\mu\nu}^{abcd} \in \mathbb{N}_{0}$, $m^{ab}_{x,\mu} \in \{0, N^{ab}_{x,\mu}\}$ and 
$\overline{m}^{ab}_{x,\mu} \in \{0, \overline{N}^{ab}_{x,\mu}\}$, with $(a,b) \in \{(2,1),(2,3),(3,1),(3,3)\}$ represent the gauge 
degrees of freedom in this formulation, while $s_{x}^{a}, k_{x,\nu}^{ab}, \overline{k}_{x,\nu}^{ab} \in \{0,1\}$ 
are the dual variables for the fermions. In order for a configuration to be admissible, fermions have to saturate the lattice and the 
gauge constraints $C_{H}[p,k,\overline{k},m,\overline{m}]$ have to be satisfied. Each configuration comes with a weight 
given by the $W$ factors in (\ref{eq:partitionQCD2}), as well as signs, given by the loop and the gauge signs. 
Also here one needs to find a partial resummation strategy in order to use the dual representation in numerical simulations.

\section{Discussion}

In this contribution we discuss our recent results for the dual representation of lattice field theories with non-abelian symmetries,
using the Abelian color cycle / Abelian color flux (ACC/ACF) approaches. We develop the method using three examples: 
The SU(2) principal chiral model with chemical potentials, as well as SU(2) and SU(3) lattice gauge theories with and without fermions.  

The key idea of the ACC/ACF approach is to write all matrix/vector products and traces in the action
with explicit sums over the color indices. The Boltzmann factor is completely factorized such that only complex numbers remain in 
the exponent. Subsequently the individual Boltzmann factors are expanded and the resulting factors can be reorganized according
to sites or links. Thus the original degrees of freedom can be integrated out with a suitable parameterization. 
The dual form of the partition function is a sum 
over closed paths on a lattice which is the product of the space-time lattice with 
an additional dimension for color. For gauge fields these loops are 
restricted to plaquettes (ACCs), while for matter fields they can form arbitrary closed paths (ACFs). These worldline 
configurations are subject to constraints, which enforce conserved combined flux through all links (gauge theories) or all sites 
(principal chiral model) for each color channel. 

For the case of the principal chiral model we could show that all weight factors in the dual representation are real and positive and 
we present first results of a Monte Carlo simulation of the worldline formulations. For the other models we currently explore
the possibility for a partial resummation strategy to overcome the complex action problem. 

The dual representations we present here are a variant of the strong coupling expansion where all expansion coefficients are known in 
closed form. We stress at this point that for QCD there are very interesting recent developments with a differently set up for  the
strong coupling expansion \cite{deForcrand:2017fky,deForcrand:2017obe,Kim:2016izx,deForcrand:2015daa,Unger:2014oga,deForcrand:2014tha,Unger:2014tua,deForcrand:2013ufa,Unger:2012jt,Fromm:2011kq,Unger:2011in,Unger:2011it} and it would be interesting
to better understand the connection between the two versions of the strong coupling series.

\vskip3mm
\noindent
\textbf{Acknowledgments:} We thank Falk Bruckmann, Philippe de Forcrand, Mario Giuliani, Yannick Meurice,
Oliver Orasch, Tin Sulejmanpasic and Wolfgang Unger 
for interesting discussions. This work is supported by the FWF DK W 1203, "Hadrons in Vacuum, Nuclei and Stars", 
and by the FWF project I 2886-N27 in cooperation with DFG TR55, ''Hadron Properties from Lattice QCD''. 
\bibliography{lattice2017}

\begin{thebibliography}{37}

\bibitem{Chandrasekharan:2008gp}
S.~Chandrasekharan, PoS \textbf{LATTICE2008}, 003 (2008), \texttt{0810.2419}

\bibitem{deForcrand:2010ys}
P.~de~Forcrand, PoS \textbf{LAT2009}, 010 (2009), \texttt{1005.0539}

\bibitem{Wolff:2010zu}
U.~Wolff, PoS \textbf{LATTICE2010}, 020 (2010), \texttt{1009.0657}

\bibitem{Gattringer:2014nxa}
C.~Gattringer, PoS \textbf{LATTICE2013}, 002 (2014), \texttt{1401.7788}

\bibitem{Gattringer:2016kco}
C.~Gattringer, K.~Langfeld, Int. J. Mod. Phys. \textbf{A31}, 1643007 (2016),
  \texttt{1603.09517}

\bibitem{Bruckmann:2014sla}
F.~Bruckmann, T.~Sulejmanpasic, Phys. Rev. \textbf{D90}, 105010 (2014),
  \texttt{1408.2229}

\bibitem{Bruckmann:2015sua}
F.~Bruckmann, C.~Gattringer, T.~Kloiber, T.~Sulejmanpasic, Phys. Lett.
  \textbf{B749}, 495 (2015), [Err.: Phys. Lett. B751 (2015) 595],
  \texttt{1507.04253}

\bibitem{Bruckmann:2015hua}
F.~Bruckmann, C.~Gattringer, T.~Kloiber, T.~Sulejmanpasic, Phys. Rev. Lett.
  \textbf{115}, 231601 (2015), \texttt{1509.05189}

\bibitem{Bruckmann:2015uhd}
T.~Kloiber, C.~Gattringer, T.~Sulejmanpasic, F.~Bruckmann, PoS
  \textbf{LATTICE2015}, 210 (2016), \texttt{1512.05482}

\bibitem{Bruckmann:2016hes}
F.~Bruckmann, C.~Gattringer, T.~Kloiber, T.~Sulejmanpasic, PoS
  \textbf{LATTICE2016}, 062 (2016), \texttt{1611.03228}

\bibitem{Bruckmann:2016txt}
F.~Bruckmann, C.~Gattringer, T.~Kloiber, T.~Sulejmanpasic, Phys. Rev.
  \textbf{D94}, 114503 (2016), \texttt{1607.02457}

\bibitem{Bruckmann:2016fuj}
F.~Bruckmann, J.~Wellnhofer, PoS \textbf{LATTICE2016}, 054 (2016),
  \texttt{1611.05643}

\bibitem{Vairinhos:2014uxa}
H.~Vairinhos, P.~de~Forcrand, JHEP \textbf{12}, 038 (2014), \texttt{1409.8442}

\bibitem{Vairinhos:2015ewa}
H.~Vairinhos, P.~de~Forcrand, PoS \textbf{CPOD2014}, 061 (2015),
  \texttt{1506.07007}

\bibitem{Rindlisbacher:2015xku}
T.~Rindlisbacher, P.~de~Forcrand, PoS \textbf{LATTICE2015}, 171 (2016),
  \texttt{1512.05684}

\bibitem{Rindlisbacher:2016cpj}
T.~Rindlisbacher, P.~de~Forcrand, Nucl. Phys. \textbf{B918}, 178 (2017),
  \texttt{1610.01435}

\bibitem{Rindlisbacher:2017dph}
T.~Rindlisbacher, P.~de~Forcrand, PoS \textbf{LATTICE2016} (2017),
  \texttt{1703.08571}

\bibitem{Wolff:2009kp}
U.~Wolff, Nucl. Phys. \textbf{B824}, 254 (2010), [Err.: Nucl. Phys. B834 (2010)
  395], \texttt{0908.0284}

\bibitem{Wolff:2010qz}
U.~Wolff, Nucl. Phys. \textbf{B832}, 520 (2010), \texttt{1001.2231}

\bibitem{Gattringer:2017hhn}
C.~Gattringer, D.~G{\"o}schl, C.~Marchis (2017), \texttt{1709.04691}

\bibitem{Orasch:2017niz}
O.~Orasch, C.~Gattringer (2017), \texttt{1708.02817}

\bibitem{Giuliani:2017mxu}
M.~Giuliani, C.~Gattringer (2017), \texttt{1702.04771}

\bibitem{Gattringer:2016lml}
C.~Gattringer, C.~Marchis, Nucl. Phys. \textbf{B916}, 627 (2017),
  \texttt{1609.00124}

\bibitem{Marchis:2016cpe}
C.~Marchis, C.~Gattringer, PoS \textbf{LATTICE2016}, 034 (2016),
  \texttt{1611.01022}

\bibitem{Bronzan:1988wa}
J.B. Bronzan, Phys. Rev. \textbf{D38}, 1994 (1988)

\bibitem{deForcrand:2017fky}
P.~de~Forcrand, W.~Unger, H.~Vairinhos (2017), \texttt{1710.00611}

\bibitem{deForcrand:2017obe}
P.~de~Forcrand, P.~Romatschke, W.~Unger, H.~Vairinhos, PoS
  \textbf{LATTICE2016}, 086 (2017), \texttt{1701.08324}

\bibitem{Kim:2016izx}
J.~Kim, W.~Unger, PoS \textbf{LATTICE2016}, 035 (2016), \texttt{1611.09120}

\bibitem{deForcrand:2015daa}
P.~de~Forcrand, O.~Philipsen, W.~Unger, PoS \textbf{CPOD2014}, 073 (2015),
  \texttt{1503.08140}

\bibitem{Unger:2014oga}
W.~Unger, PoS \textbf{LATTICE2014}, 192 (2014), \texttt{1411.4493}

\bibitem{deForcrand:2014tha}
P.~de~Forcrand, J.~Langelage, O.~Philipsen, W.~Unger, Phys. Rev. Lett.
  \textbf{113}, 152002 (2014), \texttt{1406.4397}

\bibitem{Unger:2014tua}
W.~Unger, Acta Phys. Polon. Supp. \textbf{7}, 127 (2014)

\bibitem{deForcrand:2013ufa}
P.~de~Forcrand, J.~Langelage, O.~Philipsen, W.~Unger, PoS \textbf{LATTICE2013},
  142 (2014), \texttt{1312.0589}

\bibitem{Unger:2012jt}
W.~Unger, P.~de~Forcrand, PoS \textbf{LATTICE2012}, 194 (2012),
  \texttt{1211.7322}

\bibitem{Fromm:2011kq}
M.~Fromm, J.~Langelage, O.~Philipsen, P.~de~Forcrand, W.~Unger, K.~Miura, PoS
  \textbf{LATTICE2011}, 212 (2011), \texttt{1111.4677}

\bibitem{Unger:2011in}
W.~Unger, P.~de~Forcrand, PoS \textbf{LATTICE2011}, 218 (2011),
  \texttt{1111.1434}

\bibitem{Unger:2011it}
W.~Unger, P.~de~Forcrand, J. Phys. \textbf{G38}, 124190 (2011),
  \texttt{1107.1553}

\end{thebibliography}

\end{document}